# Spruce budworm and oil price: a biophysical analogy

Luciano Celi[a,b], Claudio Della Volpe[a], Luca Pardi[b], Stefano Siboni[a]

[a] Department of Civil, Environmental and Mechanical Engineering, University of Trento – Italy
[b] Institute for Chemical and Physical Processes, National Research Council, Pisa – Italy

## 1. Why some dynamics in Economics would be similar to the Ecological ones

The behavior of complex systems is one of the most intriguing phenomena investigated by recent science; natural and artificial systems offer a wide opportunity for this kind of analysis. The energy conversion is both a process based on important physical laws and one of the most important economic sectors; the interaction between these two aspects of the energy production suggests the possibility to apply some of the approaches of the dynamic systems analysis. In particular, a phase plot, which is one of the methods to detect a correlation between quantities in a complex system, provides a good way to establish qualitative analogies between the ecological systems and the economic ones, and may shed light on the processes governing the evolution of the system.

The aim of this speech is to highlight the analogies between some peculiar characteristics of the oil production vs. price, and show in which way such characteristics are similar to some behavioral mechanisms found in Nature.

## 2. A brief history

In a previous study[1], we tried to show how a phase plot of oil production (vs. the price) has an irregular trend (random walk) with two important features that identify as inelastic the oil market (The two lines in red, in **picture 1**). The relationship, even if only qualitatively, shows two peaks upward where the oil price became very high in few years, and then rapidly decreases (green lines)[2].

This kind of "swinging behavior" recalls some typical phenomena, investigated by some theories in the domain of complex systems. In particular, this is the case of Thom's catastrophe theory[3].

## 3. Method

The behavior shown in the phase plot (picture 1) in his simpler version could be represented as follows. We can imagine having two oil wells: one at lower cost of extraction (i.e. 100 $/barrel) and one higher (200 $/barrel)[4], as schematically shown in **picture 2**.

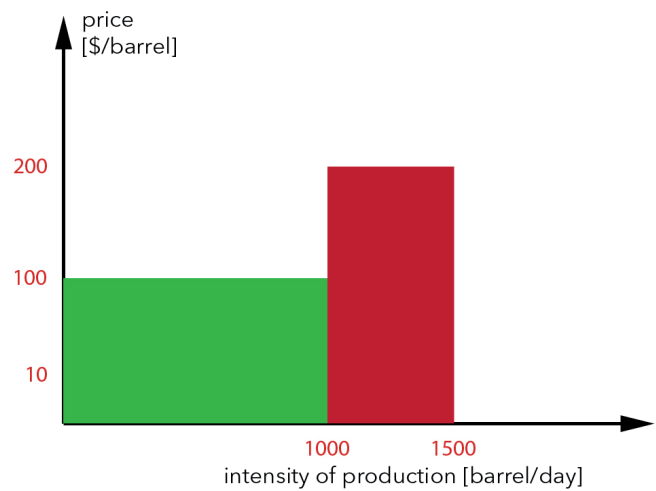

**Picture 2:** *Intensity of oil production vs. price.*

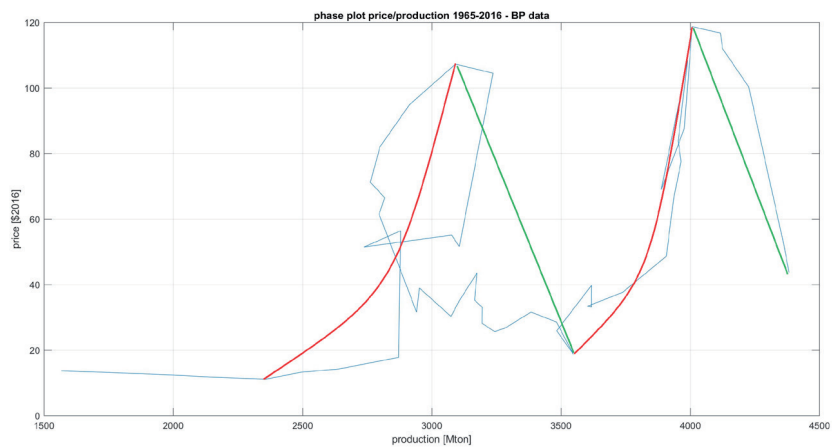

**Picture 1:** *Phase plot oil production vs. price, years 1965-2016.*

---

[1] Celi L., Della Volpe C., Pardi L., Siboni S. (2017).
[2] Obviously, we are looking at the general trend, without considering the "random walk behavior" in the middle.
[3] Thom R. (1972), cited in Scheffer M. (2009).
[4] These values are obviously arbitrary and are used only to show the behavior.

If in our hypothetical world the consumption is in the range between 0 and 1,000 barrels/day, we use the oil at a lower price (green rectangle), with a price (ideally) inside the range 0-100 $/barrel. If the intensity of consumption grows, we need to use the second stock of oil at the higher price (red rectangle). In this case, the oil price increase quite rapidly, and the phase plot should be the following, in **picture 3**.

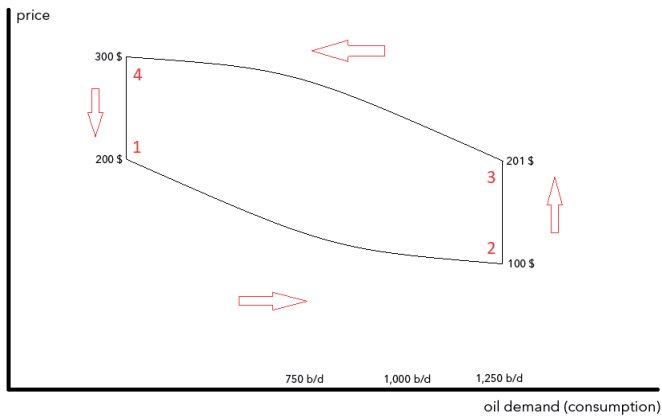

**Picture 3:** *Phase plot consumption vs. price.*

The phase plot should be like this because in our ideal world we expect that, if the oil price increases rapidly, the consumption decrease and, sooner or later, the society come back to the previous range of intensity of extraction (so in the green rectangle, picture 2).

The main characteristic of this simple oil-price dynamics, here described, is that there are two rapid movements on the cycle (rise and descent of the price: phases 4-1 and 2-3) and two slow (consumption that goes up and down, to adjust itself to the oil price: phases 1-2 and 3-4).

The analogy between this characteristic of the oil market and some dynamics in Ecology[5] is suggested at least by a model well studied[6], where we have three species in competition between them, following a variant of the Lotka-Volterra model:

1. **prey**: the American spruce, whose needles are the food of caterpillars of the species *Choristoneura Fumiferana* (this is the slow variable since the regeneration of the leaves - and not only - is a process that lasts several decades);
2. **predator**: the population of the caterpillars Choristoneura Fumiferana, considered able to vary rapidly (fast variable, since there are periodically observed demographic outbreaks of this species, considered a real scourge);
3. **"super-predator"**: the population of birds, which eat the caterpillars, but do so at a rate that we can consider constant (identified as a "natural" rate of mortality of the caterpillars themselves) because this predator actually does not feed exclusively on these caterpillars. The demographic explosion of the latter, however, saturates the space for all prey (the needles of the spruce). In this sense, we will not take into account, in the following discussion, this variable.

In the construction of the ecological model, we start with a preliminary model in which the caterpillars' population N is the only variable, while the spruces' population S is introduced as an assigned parameter. We define a range $S_- \leq S \leq S_+$ of values for S, inside the caterpillars' population could have three state of equilibrium (indicated in **picture 4**):

1. a "low" value of equilibrium $N(S)_-$, asymptotically stable;
2. an intermediate value of equilibrium $N(S)_u$, unstable, and
3. a "high" value of equilibrium $N(S)_+$, also asymptotically stable.

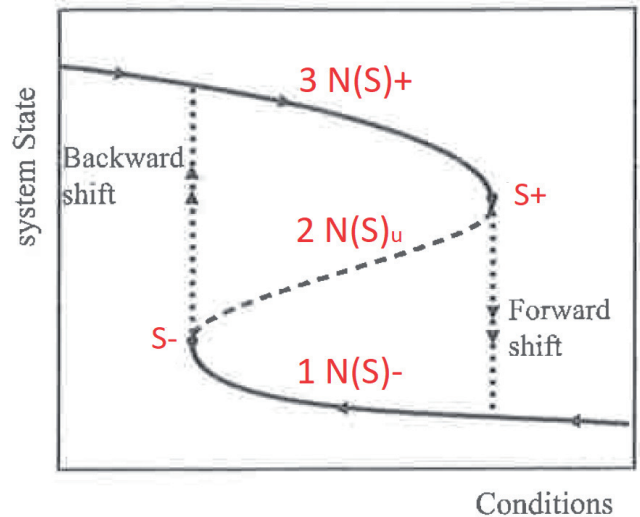

**Picture 4:** *Equilibria of the system[7].*

This notation recall us that all equilibrium states are function of the parameter S. Because of its instability, the equilibrium level $N(S)_u$ is inaccessible to the caterpillars' population, inasmuch as any variation of the population determine a rapid leaving from equilibrium and a fast convergence of the system towards $N(S)_-$ or $N(S)_+$. For $S < S_-$ only the stable equilibrium $N(S)_-$ is defined, while for $S > S_+$ we have only the stable equilibrium at $N(S)_+$. The preliminary model explains the outbreaks and collapses of caterpillars' population as imputable to slow variations imposed by the parameter S, which due to these variations "crosses" the critical values $S_-$ and $S_+$. The variations of S are assumed "almost-static", i.e. so slow to ensure that almost instantly the population of caterpillars settles at the corresponding equilibrium value (the relaxation times at equilibrium of the N population of the caterpillars are considered much shorter than those of variation of the parameter S).

The typical cycle of outbreaks and collapses of the caterpillars' population N is described as follow:

1. The process can start from a value $S < S_-$, for which the caterpillars' population corresponds to the equilibrium value $N_-$;
2. Then, we increase the value of the parameter S, corresponding to an increment of the spruce biomass. The parameter reaches and overtakes the critical value $S_-$ and still grows up until the critical value S+. In this interval, the caterpillars' population grows up but stands to the asymptotically stable equilibrium level N(S)-, because of all the possible fluctuations are lessened and reabsorbed. The compresence of the equilibrium N(S)$_u$ is not relevant, because of its instability;

---

[5] For example Kar T.K., Batabyal A. (2010); Piltzy S.H. et al. (2017).
[6] Royama T. (1984); May R.M. (1977).
[7] Scheffer M. (2009), p. 20.

3. The parameter S overtakes the critical value S+. The equilibrium N(S)- suddenly disappears, along with N(S)$_u$, and the population rapidly grows up to reach the "high" equilibrium value N(S)+, the only available and asymptotically stable. Further increments of the parameter S determine a further, but contained, increase of the caterpillars' population, in any case always corresponding to the equilibrium value N(S)+;

4. The next step is reducing the parameter S, to simulate what in reality happens: the overpopulation of caterpillars depletes the spruce biomass and determines its reduction. The parameter S reaches and overtakes the critical value S+, following its decrease until S-. The caterpillars' population slowly decreases, maintaining itself, as long as possible, close to the asymptotically stable equilibrium point N(S)+,. Also in this part of the cycle the intermediate equilibrium N(S)$_u$, even if defined, does not play any role because of its instability;

5. The last step: the parameter S finally passes below the critical value S-. This results in the destruction of the equilibrium N(S)+, as well as that of N(S)$_u$, with a consequent rapid collapse of the caterpillar population to the only available equilibrium value N(S)-. Any further decrease in S leads to a reduction in the population of the caterpillars, which however remains at the equilibrium value N(S)-;

6. Now the caterpillars' population is at the minimum level, the spruces biomass can start to grow up again, so that the parameter S grows up in turn and the cycle restarts.

The previous model with a variable (N) and a parameter (S) suggests a more complex two-variable model in which the population of caterpillars and the biomass of spruces are considered both as dynamic variables, in mutual interaction. In the further differential equation that governs the dynamics of S the characteristic constant parameters are chosen to ensure that the variation of S over time remains relatively slow. In this way, we can consider that the trends observed in the preliminary model with a single variable persist also in the new two-variable model, giving rise to a stable limit cycle characterized by two rapid growth and decrease phases of the caterpillars' population N. These rapid variations of budworm population alternate with two relatively slow growth and decrease phases of the same population, while the biomass of spruces varies always rather slowly, both increasing and decreasing, throughout the cycle.

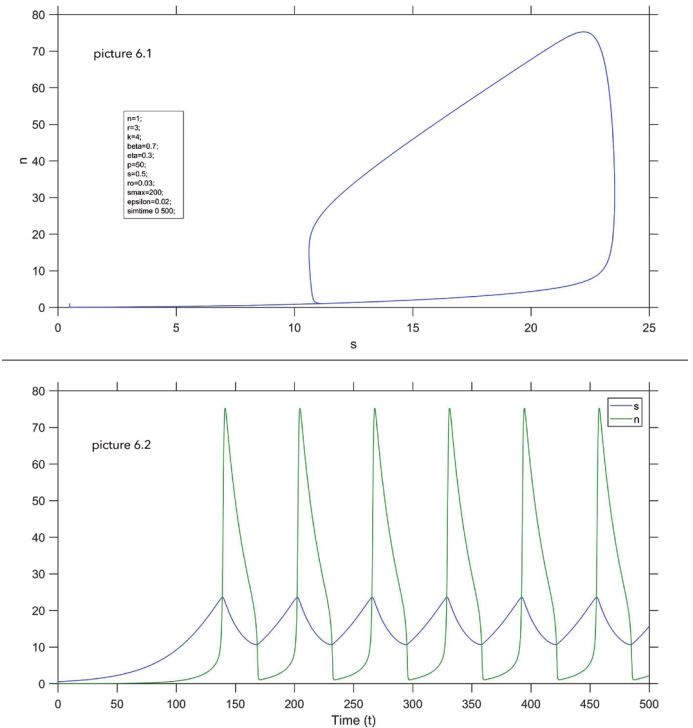

**Picture 6: 6.1**: *The Matlab simulation for the phase plot caterpillars' population vs. spruces foliage and (**6.2**) the same values in time.*

## 3.1 The transition to Economics

The shift from Ecology to the Economy (oil price-EROEI cycle) is suggested by the following qualitative considerations:

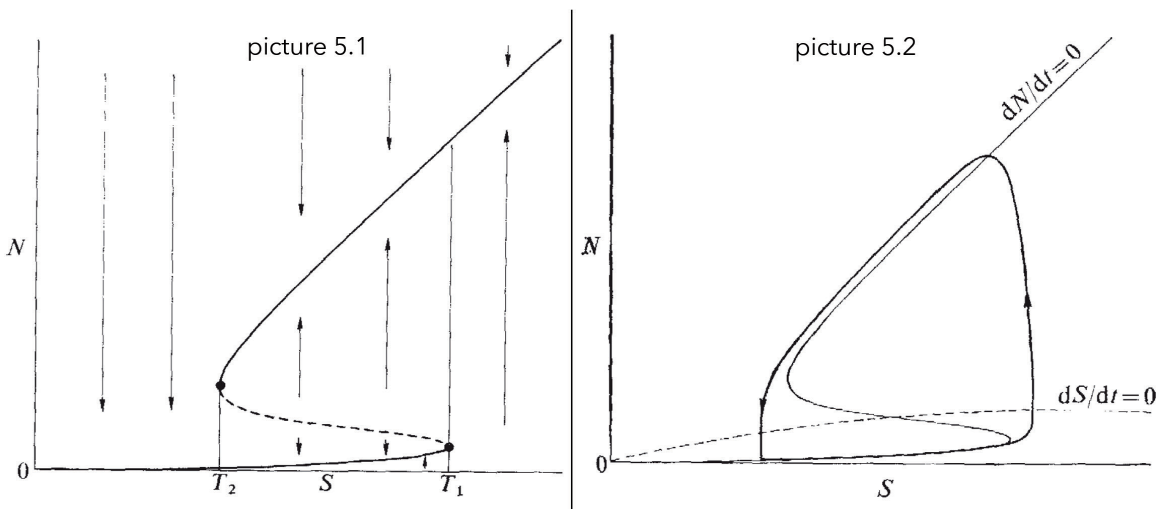

**Picture 5: 5.1**: *The equilibrium budworm population. N, shown as a function of the average leaf area per tree, S.*
**5.2**: *The (solid dN/dt=0) equilibrium curve for budworms. N, as a function of foliage S (as indicated in 5.1, where the dashed line is the non-equilibrium range N(S)$_u$)[8].*

---

[8] May R.M. (1977), picture 6 and 7 respectively.

1. The price of oil is potentially able to undergo very rapid changes, being linked to the delicate balance between supply and demand. On the contrary, EROEI presents itself as a parameter that changes slowly because its decrease naturally derives from the exploitation of deposits, while its increase can be obtained through the implementation of cultivation technologies already available (at best), the search for new deposits, or the improvement of the technologies themselves, operations that require time and significant investments;

2. A very low level of EROEI can be associated with a rapid increase in average prices, due to the difficulty of extracting the resource at low energy costs; on the other hand, a very high EROEI will favor high levels of production and a general decline in the price of the resource.

These observations suggest to identify: (a) the population N of the caterpillars with the average price of oil and (b) the biomass S of the spruces with the reciprocal of the EROEI, variable that assumes low values when the EROEI is high and vice versa. This is an index of the energy cost that must be borne to obtain a unit of useful energy.

The dynamic equations of the model spruces/*Choristoneura Fumiferana* budworm are:

$$\begin{cases} \dfrac{dN}{dt} = rN\left(1 - \dfrac{N}{kS}\right) - \beta P \dfrac{N^2}{\eta^2 S^2 + N^2} \\ \dfrac{dS}{dt} = \rho S\left(1 - \dfrac{S}{S_{max}}\right) - \varepsilon N \end{cases}$$

Where, with the above notations (N=price, S=1/EROEI), the terms in the right hand sides of the equations are:

• **rN**: it expresses the natural tendency of the price to grow up. This tendency is due to the natural limit of the resource;

• **1-N/kS**: the price actually does not increase over a limit value kS, beyond which the resource is not produced/bought. This maximum price is proportional to 1/EROEI, that is, grows with the decrease of the EROEI;

• **-βP*(N²/η²*S²+N²) = -βP*((N/S)²/(η²+(N/S)²))**: when N/S = price*EROEI (*Money Return on Energy Investment*) is high the investments are stimulated, production increases and determines a containment effect on the average price. However, the effect has a fixed limit threshold -βP. We could also imagine a growth over time of the parameter βP to account for the inevitable non-cyclical behavior of the dynamics (the resource tends, however, to run out progressively, so the cycle does not reproduce perfectly over time).

If N/S = price*EROEI is low, the financial resources available for investments are reduced and the effect on the price is minimal.

• **ρS**: the variable 1/EROEI tends to grow naturally, i.e. the EROEI naturally tends to decrease due to the exploitation of the resource;

• **1-(S/S$_{max}$)**: the variable 1/EROEI does not grow beyond a fixed maximum value S$_{max}$. In other words, EROEI does not fall below a minimum value 1/S$_{max}$, otherwise, the resource is no longer perceived as such and is not exploited. We could imagine a Smax that grows over time, always to describe the definitively non-cyclical nature of the system.

• **-εN**: high price levels have a depressive effect on 1/EROEI, that is, they favor the increase of the EROEI as they push the direct investments to a better and more efficient exploitation of the resource.

By introducing the variables E=EROEI=1/S and N=price, the equations can be rewritten as follows:

$$\begin{cases} \dfrac{dE}{dt} = \rho\left(\dfrac{1}{S_{max}} - E\right) + \varepsilon N E^2 \\ \dfrac{dN}{dt} = rN\left(1 - \dfrac{1}{k}NE\right) - \beta P \dfrac{N^2}{\eta^2/E^2 + N^2} \end{cases}$$

Putting E in abscissas and N in ordinates in the phase plane, with appropriate choices of the constant parameters we obtain the following typical limit cycle:

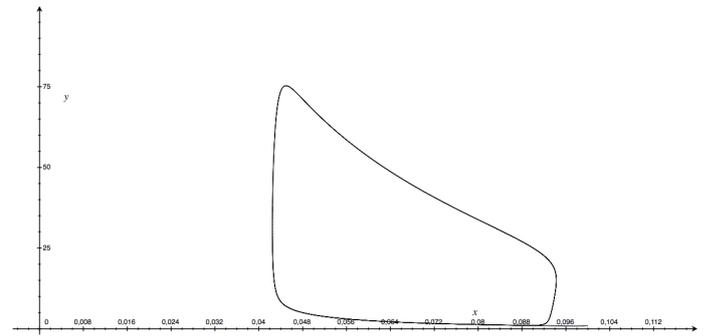

**Picture 7:** Limit cycle for N=price and S=1/Eroei.

As time goes by, the cycle is covered clockwise, with the right and left vertical sections corresponding to a sudden drop in the average price for high values of the EROEI and a rapid rise in prices when the EROEI reaches a minimum critical value, respectively. The inclined traits show an approximately hyperbolic trend: the lower one sees slowly increase the average prices to the slow decrease of the EROEI, and could correspond to the model proposed by Court and Fizaine[9], while the upper one describes a more decisive, but still relatively slow decrease of the mean price with the slow increasing of the EROEI.

## 4. Discussion

The interesting aspect of the model is that the price is not described as a function of the EROEI, because two different price levels correspond to the same value of EROEI, according to the historical phase of the economic cycle where the system is placed. Basically, we are faced with two zones of stability, one with a high content and one with a low content of "predators" (price), which are alternately reached. The reason why the sizes were chosen is that, in the oil model, the quantity that varies faster is the price, while both the total production and the EROEI (which depends on technology and investments) are sizes with too many constraints to be able to vary quickly. EROEI does not succeed because of the technical conditions of production, while production because of the constant energy hunger.

---

[9] Court V., Fizaine F. (2017).